# Cryptanalysis of an Elliptic Curve-based Signcryption Scheme[†]

Mohsen Toorani [‡]

Ali A. Beheshti

**Abstract**

*The signcryption is a relatively new cryptographic technique that is supposed to fulfill the functionalities of encryption and digital signature in a single logical step. Although several signcryption schemes are proposed over the years, some of them are proved to have security problems. In this paper, the security of Han et al.'s signcryption scheme is analyzed, and it is proved that it has many security flaws and shortcomings. Several devastating attacks are also introduced to the mentioned scheme whereby it fails all the desired and essential security attributes of a signcryption scheme.*

## 1. Introduction

The confidentiality, integrity, non-repudiation, and authentication are the most important security services in the security criteria. The encryption and digital signature are two fundamental security mechanisms that are simultaneously required in many applications. Until the previous decade, they have been viewed as important but distinct building blocks of various cryptographic systems. In public key schemes, a traditional method is to digitally sign a message then followed by an encryption (signature-then-encryption) that has two problems: Low efficiency and high cost of such summation, and the case that any arbitrary scheme cannot guarantee the security. The signcryption is a relatively new cryptographic technique that is supposed to fulfill the functionalities of digital signature and encryption in a single logical step, and can effectively decrease the computational costs and communication overheads in comparison with the traditional signature-then-encryption schemes. The first signcryption scheme was introduced by Zheng in 1997 [1] but it fails the forward secrecy of message confidentiality [2]. Zheng also proposed an elliptic curve-based signcryption scheme that saves 58% of computational and 40% of communication costs when it is compared with the traditional elliptic curve-based signature-then-encryption schemes [3]. There are also many other signcryption schemes that are proposed throughout the years, each of them having its own problems and limitations, while they are offering different level of security services and computational costs.

In a signcryption scheme, the sender usually uses the public key of recipient for deriving a session key of a symmetric encryption, while the recipient uses his private key for deriving the same session key. Exposure of session keys can be a devastating attack to a cryptosystem since such an attack typically implies that all the security guarantees are lost. In this paper, we prove that a recent signcryption scheme, i.e. Han et al.'s scheme [4] that will be referred to as HYH throughout this paper, has such vulnerability and many other security flaws. This paper is organized as follows. Section 2 briefly describes some preliminaries on signcryption and its desired attributes. Section 3 is devoted to cryptanalysis of HYH signcryption scheme, and Section 4 provides the conclusions.

## 2. Preliminaries to Signcryption

Any signcryption scheme $\Sigma = (Gen, SC, USC)$ typically consists of three algorithms: Key Generation (*Gen*), Signcryption (*SC*), and Unsigncryption (*USC*). *Gen* generates a pair of keys for any user *U*: $(SDK_U, VEK_U) \leftarrow Gen(U, \lambda)$ where $\lambda$ is the security parameter, $SDK_U$ is the private signing/decryption key of user *U*, and $VEK_U$ is his public verification/encryption key. For any message $m \in M$, the signcrypted text $\sigma$ is obtained as $\sigma \leftarrow SC(m, SDK_S, VEK_R)$ where *S* denotes the sender, and *R* is the recipient. *SC* is generally a probabilistic algorithm while *USC* is most likely to be deterministic where $m \cup \{\bot\} \leftarrow USC(\sigma, SDK_R, VEK_S)$ in which $\bot$ denotes the invalid result of unsigncryption. A formal proof for the security of signcryption is provided in [5].

Any signcryption scheme should have the following properties [6]:

1) **Correctness:** A signcryption scheme is correct only if for any sender S, recipient R, and message $m \in M$, $\exists USC(SC(m, SDK_S, VEK_R), SDK_R, VEK_S) = m$.

2) **Efficiency:** The computational costs and communication overheads of a signcryption scheme should be smaller than those of the best known signature-then-encryption schemes with the same provided functionalities.

3) **Security:** Any signcryption scheme should simultaneously fulfill the security attributes of encryption and those of a digital signature. Such properties mainly include: *Confidentiality, Unforgeability, Integrity,* and *Non-repudiation*. Some signcryption schemes provide some additional attributes such as *Public verifiability* and *Forward secrecy of message confidentiality* while the others do not provide them. Such properties are the attributes that are required in some applications while the others may not require them. *Public verifiability* is not a security attribute but it can be regarded as a facility. Hereunder, the above-mentioned attributes are briefly described.

- **Confidentiality:** It should be computationally infeasible for an adaptive attacker to gain any partial information on the contents of a signcrypted text, without knowledge of the sender's or designated recipient's private key.
- **Unforgeability:** It should be computationally infeasible for an adaptive attacker to masquerade an honest sender in creating an authentic signcrypted text that can be accepted by the unsigncryption algorithm.
- **Non-repudiation:** The recipient should have the ability to prove to a third party (e.g. a judge) that the sender has generated the signcrypted text. This ensures that the sender cannot deny his previously signcrypted texts.
- **Integrity:** The recipient should be able to verify that the received message is the original one that was signcrypted by the sender.
- **Forward Secrecy of message confidentiality:** If the long-term private key of the sender is compromised, no one should be able to extract the plaintext of previously signcrypted texts. In a regular signcryption scheme, when the long-term private key is compromised, all the previously issued signatures will not be trustworthy anymore. As the cryptographic computations are performed more frequently on poorly protected devices such as mobile phones, the threat of key exposure is becoming more acute and the forward secrecy seems an essential security attribute in such systems.
- **Public Verifiability:** Any third party can verify that the signcrypted text is the valid signcryption of its corresponding message, without any need for the private key of sender or recipient.

Many of available signcryption schemes involve modular exponentiation while some of them including the HYH signcryption scheme are based on elliptic curves. The elliptic curve-based schemes are usually based on difficulty of *Elliptic Curve Discrete Logarithm Problem* (ECDLP) that is computationally infeasible under certain circumstances [7]. The elliptic curve-based systems can attain to a desired security level with significantly smaller keys than those of required by their exponential-based counterparts. This can enhance the speed and leads to efficient use of power, bandwidth, and storage that are the basic limitations of resource-constrained devices [8].

## 3. Cryptanalysis of Han et al.'s Scheme

The signcryption and unsigncryption stages of the Han et al.'s signcryption scheme (HYH) [4] are depicted in Figure 1 where the deployed notations are described in Figure 2. The public keys of *Alice* and *Bob* are generated as $U_A = d_A G$ and $U_B = d_B G$ respectively. HYH aims to provide the attributes of *confidentiality*, *unforgeability*, *integrity*, *non-repudiation*, and *public verifiability*. However, as we prove in this section, it has several security flaws so that it fails all the desired security attributes of a signcryption scheme. Throughout this section, *Alice* is the sender, *Bob* is the designated recipient, and *Mallory* is the malicious active attacker.

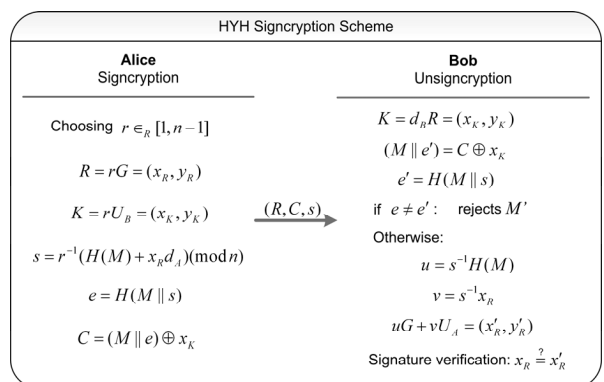

**Figure 1. HYH Signcryption Scheme [4]**

| Notations Guide | | | |
|---|---|---|---|
| $\in_R$ | Chosen randomly | $O$ | Point of elliptic curve at infinity |
| $M$ | Plaintext | $n$ | Order of $G$ ($nG=O$) |
| $C$ | Ciphertext | $x_R$ | x-coordinate value of point R |
| $s$ | Digital signature | $y_R$ | y-coordinate value of point R |
| $H$ | One-way hash function | $d_A/U_A$ | Private/Public key of *Alice* |
| $\|\|$ | Concatenation | $d_B/U_B$ | Private/Public key of *Bob* |
| $G$ | Base point of elliptic curve | | |

**Figure 2. Explanation of deployed notations**

1) The security of HYH completely depends on the secrecy of random number $r$. It does not have any resilience to disclosure of such ephemeral parameter, and the long-term private key of *Alice* $d_A$ will be simply divulged with disclosure of $r$. The point $R$ is obtained as $R = rG$ and it is clearly sent to *Bob*. If *Mallory* knows the corresponding $r$ of $R$, he can easily deduce the static private key of *Alice* from an intercepted pair of $(R, C, s)$. He calculates $K = rU_B = (x_K, y_K)$, and decrypts the ciphertext as $M \| e' = C \oplus x_k$. He then deduces the long-term private key of *Alice* as $d_A = x_R^{-1}(rs - H(M)) \bmod n$. Therefore, the confidentiality, unforgeability, non-repudiation, and other claimed security attributes of HYH completely depend on the secrecy of $r$ and will be completely failed with its disclosure. This attack is feasible due to the weak session key establishment of HYH.

Although it is believed that finding the corresponding $r$ of a specific $R$ is in deposit of solving the ECDLP, it cannot be used for concluding the claimed security attributes of HYH. Resilient to disclosure of random parameter $r$ is nowadays one of the most important and essential security attributes of any key exchange protocol so that it has been considered in many standard and secure protocols such as MQV [9] and HMQV [10] that are approved by national agencies such as NSA. However, HYH does not take benefit of such important attribute and it is completely vulnerable to disclosure of such ephemeral parameter. Although finding the corresponding $r$ of a specific $R$ is generally in deposit of solving the ECDLP, there are some practical situations where *Mallory* can defeat HYH and deduce the private key of *Alice* without any need for solving the ECDLP. Hereunder, we describe two practical scenarios for the mentioned state.

- The first feasibility is that many applications boost their performance by pre-computing the ephemeral pairs of $(r, R)$ for their later uses. This may be applied to resource-constrained devices as well as high volume servers. In this case, the stored pairs are more vulnerable to leakage than the long-term private keys. The former is typically stored on disks and hence is exposed to more vulnerability while the latter may be stored on a hardware protected storage media. If *Mallory* could have any access to such stored pairs, he can easily deduce the long-term private key of *Alice* by following the above-mentioned method.

- The second feasibility is to misuse the possible weaknesses of the deployed random number generators. The generated random numbers are actually pseudo-random and may have some biases, especially when they are generated in the resource-constrained devices. *Mallory* runs the deployed random number generator of his victims, generates the most probable pairs of $(r, R)$, and saves them offline. He then intercepts the *Bob*'s terminal that would have many transactions everyday (e.g. *Bob* can be a bank while *Alice* is a customer). *Mallory* considers the clearly sent $R$ in the intercepted messages and picks those messages for which he has their $R$ in his compiled list. He simply deduces the long-term private keys of all the corresponding senders from such chosen signcrypted texts by following the above-mentioned method. This can be considered as a *chosen-ciphertext attack*. He can use the deduced private keys for impersonating himself as the legitimate users and performing his malicious activities. If *Mallory* aims a definite entity, he may wait until his definite victim sends an $R$ that he has it in his compiled list. Until then, *Mallory* can enrich his list.

Although the mentioned attack works for awkward implementations of HYH, it is completely regarded to its weak session key derivation function that includes a simple elliptic curve point multiplication and taking the x-coordinate of the product as the session key.

2) An extra chosen-ciphertext attack is also applicable to HYH since it uses a simple XOR for the encryption. The chosen-ciphertext security (IND-CCA) is a standard and acceptable notion of security for a public key encryption scheme [11]. The chosen-ciphertext attack to HYH can be accomplished by choosing those texts that are signcrypted with the same random number $r$ and consequently having the same clearly sent value of $R$. For such chosen ciphertexts, we have:

$$C_1 \oplus C_2 = (M_1 \oplus M_2) \| [H(M_1 \| r^{-1}(H(M_1) + x_R d_A)) \\ \oplus H(M_2 \| r^{-1}(H(M_2) + x_R d_A))] \tag{1}$$

Where $C_1$ and $C_2$ are the corresponding ciphertexts of messages $M_1$ and $M_2$ respectively. Expression (1) shows a linear relationship between the plaintext and ciphertext that can be a subject for several cryptanalysis methods such as linear cryptanalysis.

3) In certificate-based public key schemes, after doing the certificate validation, the validity of public keys should be verified using the validated certificates. Otherwise, the certificates and public keys can be easily forged and the scheme will succumb to the *man-in-the-middle* attack. The process of certificate validation includes [12]:
(a) Verifying the integrity and authenticity of the certificate by verifying the CA's signature on the certificate.
(b) Verifying that the certificate is not expired.
(c) Verifying that the certificate is not revoked.
However, HYH does not consider such considerations.

4) HYH does not consider the public key validation so it is feasible to get certificates for the invalid public keys. An invalid public key is of a small order resided on an invalid-curve that can be misused for an *invalid-curve attack* [7]. The public key of user *U*, $U_U = (x_{U_U}, y_{U_U})$ is valid if all the following conditions are simultaneously satisfied [13]:
(a) $U_U \neq O$.
(b) $x_{U_U}$ and $y_{U_U}$ should have the proper format of $F_q$ elements.
(c) $U_U$ should satisfy the defining equation of *E*.

Traditionally, the public key validation is not considered in the PKI standards (such as [14] and [15]), and the *Certificate Authority* (*CA*) just performs a proof of possession by checking the user's signature over a message of a predetermined format so it is feasible to get a certificate for an invalid public key if the public key validation is not considered. Antipa et al. [13] demonstrated how to get a certificate for an invalid public key when *CA* uses the ECDSA. In HYH, the *CA* does not verify whether each entity really possesses the corresponding private key of its claimed public key or not. Such shortcoming exposes it to the mentioned vulnerability.

5) The delivery confirmation or a receipt from the recipient is necessary for some applications. Although HYH is a one-pass scheme, the implementer may add a confirmation step in which *Bob* sends *Alice* a confirmation message perhaps in addition to a *Message Authentication Code* (MAC) in which the session key of encryption is used as the key. Since the validity verification of ephemeral public key (i.e. the point *R*) is not included in the unsigncryption phase of the HYH, it can be misused for an *invalid-curve attack* [13] whereby *Alice* is capable of deducing the long-term private key of *Bob*. Here is how the attack works. *Alice* chooses an invalid-curve containing a point $W_i$ of small order $g_i$. She uses $W_i$ instead of *R*, proceeds the signcryption, and sends $(W_i, C, s)$ to *Bob*. Consequently, *Bob* computes $K = d_B W_i = (x_K, y_K)$ and performs the unsigncryption. Finally, when *Bob* sends the confirmation message *M'* and its corresponding tag $z = MAC_{x_K}(M')$ to *Alice*, due to the small order of point $W_i$, *Alice* can easily determine a point $K' \in <W_i>$ satisfying $z = MAC_{x_{K'}}(M')$. Hence, with $\frac{g_i}{2}$ number of trials, *Alice* can find $d_{g_i}^2 \equiv d_B^2$. She selects other $W_i$ points of different orders $g_i$, and repeats the above-mentioned procedure. The orders of selected $W_i$ points should be relatively prime so we should have $\gcd(g_i, g_j) = 1$, $\forall i \neq j$. Such points can be selected from different invalid-curves. Each round of attack gives $d_{g_i}^2 \equiv d_B^2$. Ultimately, *Alice* finds the private key of *Bob* using the *Chinese Remainder Theorem* (CRT) [12] while *Bob* is unaware that such an attack is taking place.

6) HYH is vulnerable to the *Unknown Key-Share* (UKS) attack. In an UKS attack [16], two parties compute the same session key but have different views of their peers in the key exchange. In an UKS attack, an adversary interferes with *Alice*'s and *Bob*'s communication so that *Alice* correctly believes that her session key is shared with *Bob*, while *Bob* mistakenly believes that the session key is shared with another entity. This can be accomplished whenever *Mallory* can convince one of the honest parties that he has the knowledge of the session key. Further issues on the practical attack scenarios and the significance of the UKS attack is provided in [17]. The UKS attack is feasible when a key exchange protocol fails to provide an authenticated binding between the session key and identifiers of the honest entities. Since the private key and identifier of *Alice* are not involved in the session key derivation function of the HYH, it does not have any resilience to the UKS attack.

7) Domain parameters of HYH are not exactly selected. Practically, there are some considerations that should be taken into account in selecting the domain parameters of elliptic curves, in order to thwart several potential attacks to elliptic curve-based schemes [18]. Such considerations are not considered in domain parameters' specifications of the HYH that can make it vulnerable to several kinds of attacks if the implementer unconsciously selects the domain parameters in the range of such non-stated conditions. Indeed, to thwart the *small subgroup attacks* [7], the point *G* should be of prime order $n$ and we should have $n > 4\sqrt{q}$ [9] but they are not considered in HYH that can make it vulnerable to the *small subgroup attack*. Furthermore, to protect against other known attacks to special classes of elliptic curves, $n$ should not divide $q^i - 1$ for all $1 \leq i \leq f$ ($f = 20$ suffices in practice [19]), $n \neq q$ should be satisfied, and the curve should be non-supersingular [9]. Such considerations are not also considered in HYH.

8) There is not any provision for the key control in HYH so the plaintext may be encrypted with a weak or even a full-zero key. There is also no checking for $K \neq O$.

9) Although it is not claimed in [4] that HYH provides the *forward secrecy of message confidentiality*, we found it noteworthy to specify that HYH does not provide such an attribute. The outsider and insider security are two notions of security that are usually considered in the signcryption. While the outsider security assumes that the adversary is neither sender nor the recipient, the insider security allows the adversary to be sender or recipient. The *forward secrecy of message confidentiality* is an attribute that is provided through the insider security. One may think that HYH provides such an attribute since its message confidentiality relies on two secret factors: the long-term private key of *Alice* ($d_A$), and the ephemeral random number $r$. However, anyone who has $d_B$ can simply recover the signcrypted text and deduce the corresponding random number $r$ as $r = s^{-1}(H(M) + x_R d_A) \mod n$. When $d_A$ is revealed, *Mallory* who could obtain $d_A$ may also request *Bob* to compute the corresponding $r$ for him so he can simply recover the signcrypted text without any need for knowledge of $d_B$. Regardless of its practical benefits, this invalidates the definition of *forward secrecy of message confidentiality*.

## 4. Conclusions

The security of Han et al.'s signcryption scheme [4] is analyzed in this paper, and it is proved that it has many security flaws. Several devastating attacks are also introduced to the mentioned scheme whereby it fails all the desired and essential security attributes of a signcryption scheme. It is proved that the most important security vulnerability of Han et al.'s scheme is due to its weak session key establishment while it encrypts messages by a simple XOR, and the case that it does not consider many essential considerations that should be taken into account in elliptic curve and public key cryptography. There are also other shortcomings that were explained throughout the paper.